# Observed lifespan differential – global trends, policy impact and computational methods

Toni Ćosić[1] · Roko Mišetić[2] · Hrvoje Štefančić[3]

**Abstract**

The issue of longevity has been long time recognized as one of key concepts in demography. A particular aspect of longevity addressed in this paper is the difference in average observed duration of life for female and male population, called observed lifespan differential. Using the data from Human Mortality Database, the dynamics of the observed lifespan differential is studied for a large number of countries worldwide from 1960 to 2014. An interesting phenomenon that the growing trend of the observed lifespan differential at the beginning of the studied interval does not persist, i.e. that it reverts to stagnation or even decline is revealed for a large majority of countries in the dataset. In a number of case studies, a strong association of lifespan dynamic with disruptive events such as wars, dissolutions or integrations of states or policy measures is demonstrated. Finally, a novel method of calculating the observed lifespan differential of a population from mortality indicators of its subpopulations is introduced and applied to the analysis of the observed lifespan differential in Israel from 1990 to 2000.

**Introduction**

The phenomenon of longevity is present as a subject of many scientific works due to its connection with various important aspects of social life such as, for example, the quality of life and life conditions. Thus, it is often emphasized that "length of life is a fundamental dimension of human prosperity." (Tuljapurkar & Edwards 2011, 498).

Longer lifespan is considered as one of the fundamental civilisation values and also as an indicator of wellbeing. It should be noted that the study of this phenomenon has been continually present in the world for the past 200 years (Aburto et al. 2018; Bergeron-Boucher et al. 2015; Edwards and Tuljapurkar 2005; Gillespie et al. 2014; Hart and Hertz 1944; Seaman et al. 2015; van Raalte 2011; Vaupel et al. 2011). In general, longevity is caused by two groups of factors: internal factors which are biologically/genetically conditioned, and external factors which are socially, social-psychologically and culturally conditioned (Carey 2003; Edwards and Tuljapurkar 2005; Gjonça et al. 2005; Kalben 2002). However, the impact of these factors may be more accurately described through their synergy, rather than analysing the individual impact of either group. Although both groups act in mutual interaction, within this paper their influences will be analysed separately, in order to highlight the specifics of each of them and then, to focus on the socio-cultural aspect of the lifespan.

---

[1] Department of Sociology, Catholic University of Croatia, Ilica 242, 10000 Zagreb, Croatia (toni.cosic@unicath.hr)
[2] Department of Sociology, Catholic University of Croatia, Ilica 242, 10000 Zagreb, Croatia (roko.misetic@unicath.hr)
[3] Department of Psychology, Catholic University of Croatia, Ilica 242, 10000 Zagreb, Croatia (hrvoje.stefancic@unicath.hr)

Despite the fact that internal factors continuously influence longevity, it could be said that the impact of the external factors is often stronger and more diverse. First of all, it is related to the great influence of the development of medicine and health services. In other words, public health as well as changes in the types of work activities, work conditions and technological innovations significantly affect the decrease of mortality, especially in the younger working population (Carey 2003; Gjonça et al. 2005; Hart and Hertz 1944). Research has shown a clear positive correlation between the level of education and longevity (Sasson 2016; van Raalte 2011; van Raalte et al. 2011; van Raalte et al. 2012; van Raalte et al. 2018). There is a similar connection between socio-economic status and longevity (Gjonça et al. 2005; Jackson 1994; Sasson 2016; van Raalte 2011; van Raalte et al. 2011; van Raalte et al. 2012; van Raalte et al. 2014; van Raalte et al. 2018).

Research of change of longevity in a population is usually carried out through analysis of two indicators: average longevity and life expectancy. Both indicators have shown strong growth since the mid of 19th century to the present (Bergeron-Boucher et al. 2015; Edwards and Tuljapurkar 2005; Hart and Hertz 1944; van Raalte 2011; Vaupel et al. 2011). Except for this characteristic, research has shown a clear difference between male and female populations with respect to longevity, in the sense that women on average live longer. This feature is also the result of the impact of two mentioned factors (internal as well as external) that are described in the literature. Biologically, males and females, due to different genetic as well as hormonal structure, suffer from different diseases (Austad and Fischer 2016; Kalben 2002). It should be added that men are more susceptible to diseases that increase their mortality in their earlier years (Kalben 2002). Differences in longevity and life expectancy of men and women are greatly determined by habits, in other words, it mostly relies on lifestyle and, in the broad sense depends on cultural specificity. While a male population is more inclined to bad habits (alcohol consumption, smoking, drug abuse, etc.), women generally take more care of their health and proper nutrition (Edwards and Tuljapurkar 2005; Gjonça et al. 2005; Kalben 2002; Luy and Minagawa 2014). Furthermore, research has shown that the distribution of employment in various jobs has also had a major impact on the differences in the longevity of female and male populations. Namely, males traditionally work in a more dangerous working environment, do more damaging and stressful jobs, which in combination with their biological predispositions results in an average reduction in life expectancy (Gjonça et al. 2005; Kalben 2002).

The understanding of differential in lifespan between women and men is important for many reasons. It reflects some of the most important differences between sexes, namely, how much longer do women on average live longer than men. Any changes in the observed lifespan differential reflect differences of average longevity of males and females. In the scientific literature, it is usual to accept the difference in the life expectancy or in the lifespan of males and females, as a proven and verified fact. Many scientific works report and study rising trends of measures of longevity (duration of life) in particular of lifespan and life expectancy. The sex differences or ratios of life expectancy, mortality and morbidity have also been studied and their explanation in terms of biological, behavioural and other social factors, including methodological challenges in reporting data have been proposed (Waldron 1993; Case and Paxson 2005; Preston and Wang 2006; Oksuzyan et al. 2008; Rogers et al. 2010). However, so far there has been a significant lack of analysis of the observed lifespan differential. Therefore, the central focus of this paper is on the difference of average observed lifespan between female and male populations in various countries over the world. The observed lifespan differential through the long period shows the connection with external social factors such as wars but also healthcare and social policies. On that basis, one can assume that the dynamic of changes in the observed lifespan differential can be regarded as an indicator of various social influences and processes. The observed lifespan differential is a practically important quantity since it reflects realized difference in the longevity of females and males.

The observed average lifespan exhibits almost monotonous growth both for female and for male populations in virtually all countries in the world. As mentioned before, this fact is considered to be a hallmark of universal progress in the quality of human life globally. However, an equally important question is if the observed lifespan differential, as a measure of sex difference, also exhibits such a universal unidirectional behaviour. Indeed, there is no simple a priori argument whether the observed lifespan differential is expected to grow, decline or stagnate. The main research goal of this paper is establishing the pattern of the observed lifespan differential dynamics for various countries worldwide. Specific research goals of this paper encompass:

- identifying possible universal trends in the observed lifespan differential in the studied dataset;
- providing case studies of strong association of significant changes in the observed lifespan differential with disruptive events such as wars or dissolutions and integrations of states;
- introducing methods, both exact and approximate, for calculating the observed lifespan differential of a population from the mortality indicators of its subpopulations.

Since the observed lifespan differential is dependent on age-sex structure of the studied population, we do not make detailed quantitative comparisons of the observed lifespan differentials for different populations, but present (semi)quantitative similarities in their multidecadal trends.

The structure of the paper is the following. After the introduction given in the first section, in the second section titled Methods and data the observed lifespan differential is defined and the method for its calculation is elaborated. This section also brings the description of the dataset from the Human Mortality Database used in this paper. The third section, titled Results, brings several subsections in which general trends of the observed lifespan differential are surveyed and a number of case studies on lifespan differential and its connection to disruptive events is presented. A subsection is devoted to the precision analysis of the computational method using subpopulations. The final subsection is dedicated to the application of the developed method to the observed lifespan dynamics in Israel from 1990 to 2000. Five year moving averages of the observed lifespan differential data for all countries from the dataset are depicted in the Appendix.

**Methods and data**

In the first subsection of this section, titled Lifespan differential, the concept of the observed lifespan differential is introduced and rigorously mathematically defined. A formula for the observed lifespan differential of a population in terms of mortality indicators of its subpopulations is derived. The second subsection, titled Data, brings the description of the used dataset from the Human Mortality Database.

*Observed lifespan differential*

The central quantity of interest in this paper is the female-male observed lifespan differential. In the remainder of this section we provide precise technical definition for the said observed lifespan differential and propose an efficient approximate computational scheme for the observed lifespan differential of a population consisting of subpopulations with different mortality properties. In this paper we restrict our analysis to populations consisting of two non-overlapping subpopulations.

Let us consider a population for which sex-age specific mortality data is available. We consider a total population, denoted by a superscript *tot*, and its two subpopulations, denoted by superscripts *1* and *2*. Each of these (sub)populations is further divided into female and male subpopulations, denoted by subscripts *f* and *m*, respectively.

The quantities describing the mortality are:

$d_m^j(i,t)$ - number of males of age between *i* and *i+1* in population *j* dying in year *t*;

$d_f^j(i,t)$ - number of females of age between *i* and *i+1* in population *j* dying in year *t*;

$D_m^j(t) = \sum_{i=0}^{\infty} d_m^j(i,t)$ – total number of males in population *j* dying in year *t*;

$D_f^j(t) = \sum_{i=0}^{\infty} d_f^j(i,t)$ – total number of females in population *j* dying in year *t*.

As a practical step, in the definitions of sex-specific total numbers of deaths we set the upper limit of summation to infinity. This step simplifies notation, especially in cases when the maximal age for which data are recorded changes over time. Average observed lifespan values for males and females are defined as:

$\bar{l}_m^j(t) = \frac{\sum_{i=0}^{\infty}(i+1/2)d_m^j(i,t)}{D_m^j(t)}$ – average observed lifespan (age at dying) of male population *j* dying in year *t*;

$\bar{l}_f^j(t) = \frac{\sum_{i=0}^{\infty}(i+1/2)d_f^j(i,t)}{D_f^j(t)}$ – average observed lifespan (age at dying) of female population *j* dying in year *t*.

Finally, the female-male observed lifespan differential is defined as a difference of the average observed lifespan of females and average lifespan of males:

$\Delta l^j(t) = \bar{l}_f^j(t) - \bar{l}_m^j(t)$ – observed lifespan differential of population $j$ in year $t$.

The relations of corresponding quantities for the total population and its non-overlapping subpopulations are:

$$d_m^{tot}(i,t) = d_m^1(i,t) + d_m^2(i,t), \quad (1)$$

$$d_f^{tot}(i,t) = d_f^1(i,t) + d_f^2(i,t), \quad (2)$$

$$D_m^{tot}(t) = D_m^1(t) + D_m^2(t), \quad (3)$$

$$D_f^{tot}(t) = D_f^1(t) + D_f^2(t). \quad (4)$$

We are interested in relation between the observed lifespan differential for the total population and observed lifespan differentials of its two subpopulations. After some algebra, it is straightforward to show

$$\Delta l^{tot}(t) = \frac{D_f^1(t)\Delta l^1(t) + D_f^2(t)\Delta l^2(t)}{D_f^{tot}(t)} + \left[\frac{D_f^1(t)}{D_f^{tot}(t)} - \frac{D_m^1(t)}{D_m^{tot}(t)}\right]\left(\bar{l}_m^1(t) - \bar{l}_m^2(t)\right). \quad (5)$$

This expression can be conveniently written as

$$\Delta l^{tot}(t) = \Delta l_{app}^{tot}(t) + \Delta l_{corr}^{tot}(t), \quad (6)$$

where

$$\Delta l_{app}^{tot}(t) = \frac{D_f^1(t)\Delta l^1(t) + D_f^2(t)\Delta l^2(t)}{D_f^{tot}(t)}, \quad (7)$$

$$\Delta l_{corr}^{tot}(t) = \left[\frac{D_f^1(t)}{D_f^{tot}(t)} - \frac{D_m^1(t)}{D_m^{tot}(t)}\right]\left(\bar{l}_m^1(t) - \bar{l}_m^2(t)\right). \quad (8)$$

This form of writing down the expression for $\Delta l^{tot}(t)$ is useful as it represents it as a sum of two terms of which the first term, $\Delta l_{app}^{tot}(t)$, is the leading contribution and the second term, $\Delta l_{corr}^{tot}(t)$, represents a small correction. A simple argument for such a hierarchy of these two terms can be provided.

The first term, given in (7), is a weighted average of the observed lifespan differentials for subpopulations 1 and 2, with the corresponding total numbers of female deaths as weights. By definition, this contribution is between $\Delta l^1(t)$ and $\Delta l^2(t)$ and it is, therefore of the order of the typical observed lifespan differential (roughly between 5 and 15 years). The second term is a product of two factors. The first factor, $\left[\frac{D_f^1(t)}{D_f^{tot}(t)} - \frac{D_m^1(t)}{D_m^{tot}(t)}\right]$, is the difference of proportion of total number of female deaths of subpopulation 1 in the total number of female deaths in the entire population and its counterpart for male subpopulation. This factor is expected to be of order of several percent or less. On the other hand, the second factor, $(\bar{l}_m^1(t) - \bar{l}_m^2(t))$, is of the order of the observed lifespan differentials. This semi-quantitative argumentation leads us to the conclusion that we should expect the second term to be a correction at the percent level or less. This finding is numerically verified in the section Results.

*Data*

In this paper we use the data from Human Mortality Database (University of California, Berkeley (USA) and Max Planck Institute for Demographic Research (Germany)) which are available at www.mortality.org or www.humanmortality.de. This database contains yearly mortality data for 40 countries worldwide. The range of available yearly data varies from country to country. In this paper we use data from 1960 to 2014 whenever available. The mortality figures are further segmented according to age and sex. In general, age segmentation is performed in yearly intervals, going from 0 to 110+. For practical reasons, we approximate the age of 110+ category to be 110. Given very small numbers in this category for all countries, this approximation makes a negligible effect on results.

**Results**

In this section we present results for the observed lifespan differential for all countries of the studied dataset. The most important result reported in this paper is the appearance of reversal of trend in the observed lifespan differential for almost all countries in the studied dataset. Some hypotheses for further study of this important feature are outlined in the section Discussion and directions for future research.

The presentation of results in this section is organized as follows: First we present some general trends and patterns observed for most of countries in the studied dataset and analyze some of their most prominent properties; then we display several examples in which the observed trends can be associated with disruptive events such as wars, dissolutions of states or accession of countries to supranational unions; finally in the last part of this section we use our method of calculating the lifespan differential for the total population from mortality parameters of its subpopulations and apply it to the study of impact of immigration to dynamics of lifespan differential for Israel in period from 1990 to 2000.

*Global trends*

In Figure 1 the movement of the observed lifespan differential in the period of several decades is presented for ten countries from the available dataset from Human Mortality Database where the observed behavior is clearly visible. A prominent feature in the observed lifespan differential trend is a change from a growing trend to a decreasing (or stagnating) trend. The timing of this change varies from country to country, possibly connected to different phases of societal development in these countries. Given that the values of observed lifespan differential differ considerably from country to country, it is even more interesting to observe this almost universal pattern. As for some countries with smaller populations the effect of stochastic fluctuations is significant, in the Appendix we present the observed lifespan differential dynamics for all countries in the studied dataset as a five-year moving average. The mentioned change of trend is visible in a large majority of studied countries.

The main goal of this paper is to present and stress the existence of such an almost universal behavior of the observed lifespan differential. We discuss some possible sources of this regime change in the section Discussion and directions for future research. However, in this paper we do not undertake full statistical analysis of sources of such dynamics which is left for future work. Finally, despite the presence of the observed change of trend in a large majority of studied countries, there are some notable examples such as Japan, Norway and Slovakia where there is no indication of the change from the growing trend and peculiar case of Iceland where there is a, although fluctuating, still persistent downward trend of the observed lifespan differential.

A cautionary remark is in order. It is hard to speak of universality without detailed statistical analysis. Clearly there is some differentiation between all countries in the dataset (e.g. countries in Figure 1 have a prominent peak, some others exhibit a combination of initial growth and subsequent stagnation of the lifespan differential). When we speak of universality, we claim that there is a following almost universal property: the growing initial trend in approximately 1960-1980 is not continued (there is significant and visible declination from it) in the years 1990-2014. In other words, there is a reversal of trend – from the growth of the observed lifespan differential to stagnation or decline.

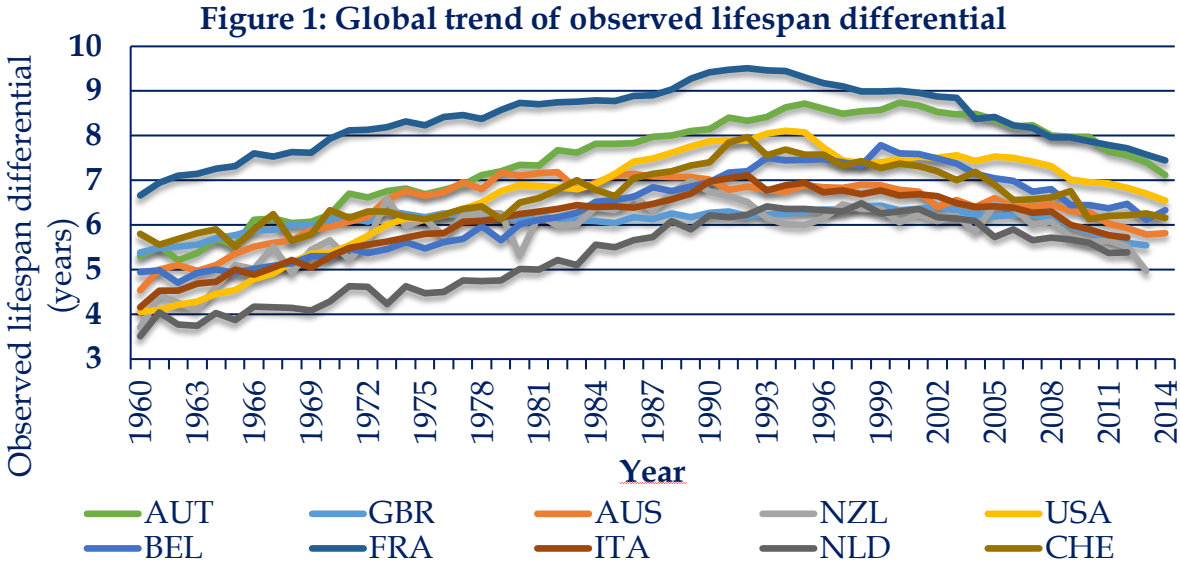

*Wartime losses*

In Figure 2 we present the observed lifespan differential for Croatia in the period from 1960 to 2014. The most outstanding detail in this figure is a prominent peak in the lifespan differential during wartime in Croatia (Homeland war), especially in years 1991 and 1992. Prior to peak, the observed lifespan differential exhibits growing trend, while in the years after the peak there is a decline in lifespan differential. The peak represents an increase of around 35 % compared to years immediately preceding or immediately following the peak years. An explanation lies in the casualties of the combatant population which was predominantly younger or middle-aged male population. Indeed, the graph of the average observed lifespan for male population exhibits a visible dip in years 1991 and 1992. The civilian (non-combatant) casualties are expected to be more evenly distributed among male and female population and different ages and their contributions to the lifespan differential are expected to largely cancel.

These results point at lifespan differential as a potentially sensitive indicator of importance of wartime losses for the total population. Namely, the prominence of the peak (measured by the height of the peak relative to adjacent years) is large only if the numbers of male deaths at younger age is significantly increased. A more systematic analysis of this proposal is left for future work.

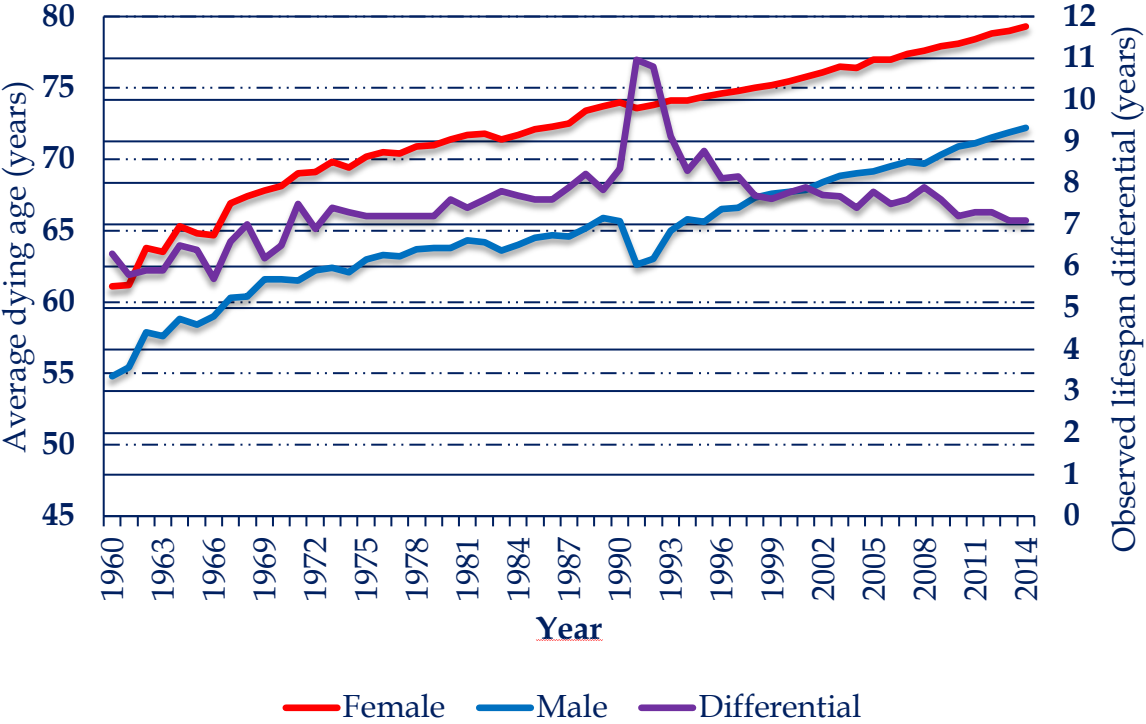

**Figure 2: Effect of wartime losses - Croatia**

*Influence of policy and regulation*

In this subsection we present several case studies which illustrate influence of disruptive phenomena, such as regulatory changes or substantial changes in immigration volume, on the observed lifespan differential dynamics. These case studies provide indication what some of forces driving the observed lifespan differential dynamics might be. In the first three case studies we present peculiarities of the observed lifespan differential dynamics that can be associated with regulatory changes, in the fourth case study we perform numerical tests of the approximation scheme proposed in section Methods and data, whereas in the fifth case study we consider interplay of immigration and the observed lifespan differential for the case of Israel in the period from 1990 to 2000.

Policy impact – republics of former Soviet Union

For the six former Soviet Union countries within the studied dataset (Russia, Ukraine, Belarus, Lithuania, Latvia and Estonia) there is a distinctive fall in the observed lifespan differential in the middle of 1980s, as depicted in Figure 3. A particularly interesting feature of Figure 3 is coincidence of this fall for all six countries in the same period, despite different behavior of corresponding lines in the period after gaining independence. i.e. dissolution of Soviet Union. Works in the literature (Bobadilla et al. 1997) lend credibility to hypothesis that the observed and coordinated fall in the value of lifespan differential across various Soviet Union republics could be attributed to policy measures against alcoholism (so called Anti-alcohol campaign) which were implemented throughout Soviet Union in mid 1980s. The decrease of the observed lifespan differential can be attributed to the fact that alcoholism contributes more to the mortality of male than female population, especially at younger and middle age.

This example illustrates the intensity of impact that policy measures can have on the lifespan differential.

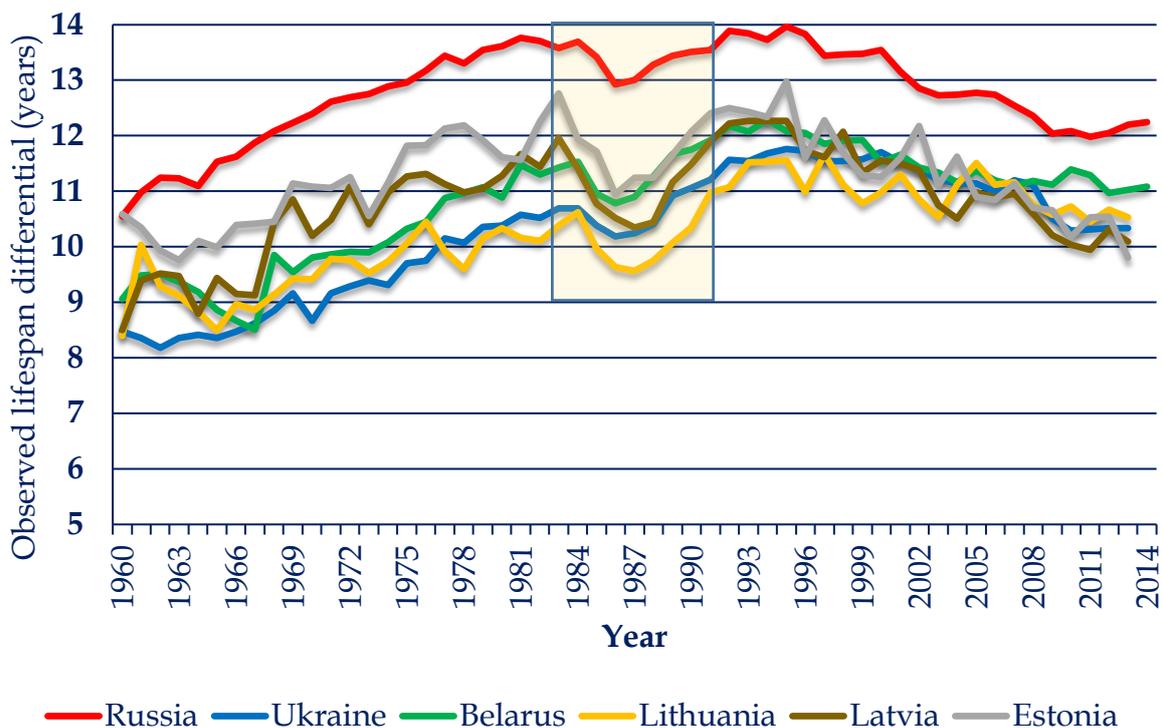

**Figure 3: Policy impact - republics of former Soviet Union**

Supranational Integration – accession of Portugal and Spain to EEC

A comparative analysis of lifespan differential for Portugal and Spain, depicted in Figure 4, makes our second example. An important moment in the studied period from 1960 to 2014 is the moment of accession of both countries to European Union (European Economic Community (EEC) at the time) in 1986. It is quite indicative that exactly around this year the difference of the observed lifespan differentials between Portugal and Spain has undergone a significant change in behavior. Namely, in the period before 1986 the lifespan differential of Portugal is significantly larger than the lifespan differential of Spain. The difference of lifespan differentials for these two countries changes sign around 1986 and in the subsequent period the difference between lifespan differential for Portugal and lifespan differential for Spain is negative and much smaller in absolute value than in the period prior to 1986, as evident from Figure 4. A natural question is if the regulatory and administrative change of accession to the EEC could be associated with the convergent trend of lifespan differentials for Portugal and Spain. It remains to be studied into more detail to which extent increased funding and accelerated development, as well as harmonization of legislature and procedures with other EEC member countries could lead to increasing proximity of lifespan differentials of Portugal and Spain.

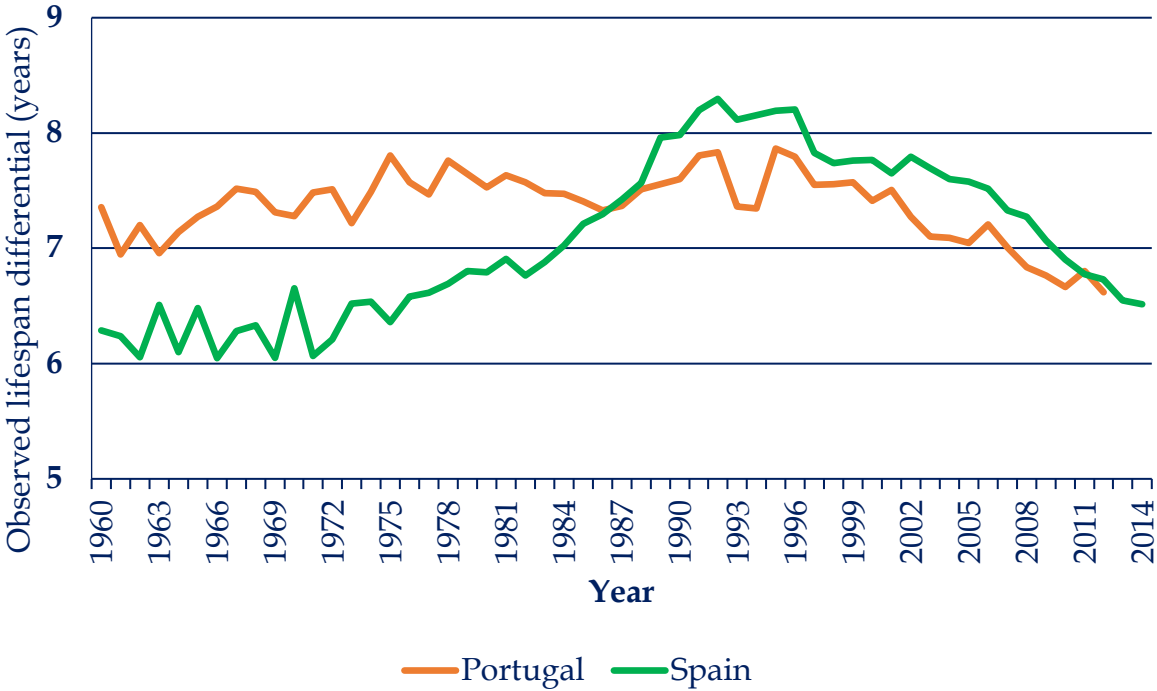

**Figure 4: Effects of supranational integration - accession of Portugal and Spain to EEC**

Dissolution of states - Czechia and Slovakia

Our next example refers to comparison of lifespan differential for Czechia and Slovakia before and after dissolution of Czechoslovakia. In Figure 5 we present the observed lifespan differential dynamics in the period from 1960 to 2014. The lifespan differential data presented in this Figure show that whereas there is very close agreement of lifespan differential in the period from 1970 to 1990, after approximately 1998 there appears a divergent trend of lifespan differential graphs for Czechia and Slovakia. It is intriguing to observe this property of the observed lifespan differential for these two countries from the perspective that before 1992 these two countries were parts of a single country in which the same set of policies were applied throughout the entire country, whereas after 1992 these two countries started implementation of their separate policies affecting the observed lifespan differential. At this level, this intriguing association is only a plausible hypothesis that different policies lead to different observed lifespan patterns. Clearly, the verification or rejection of this hypothesis requires a more dedicated study.

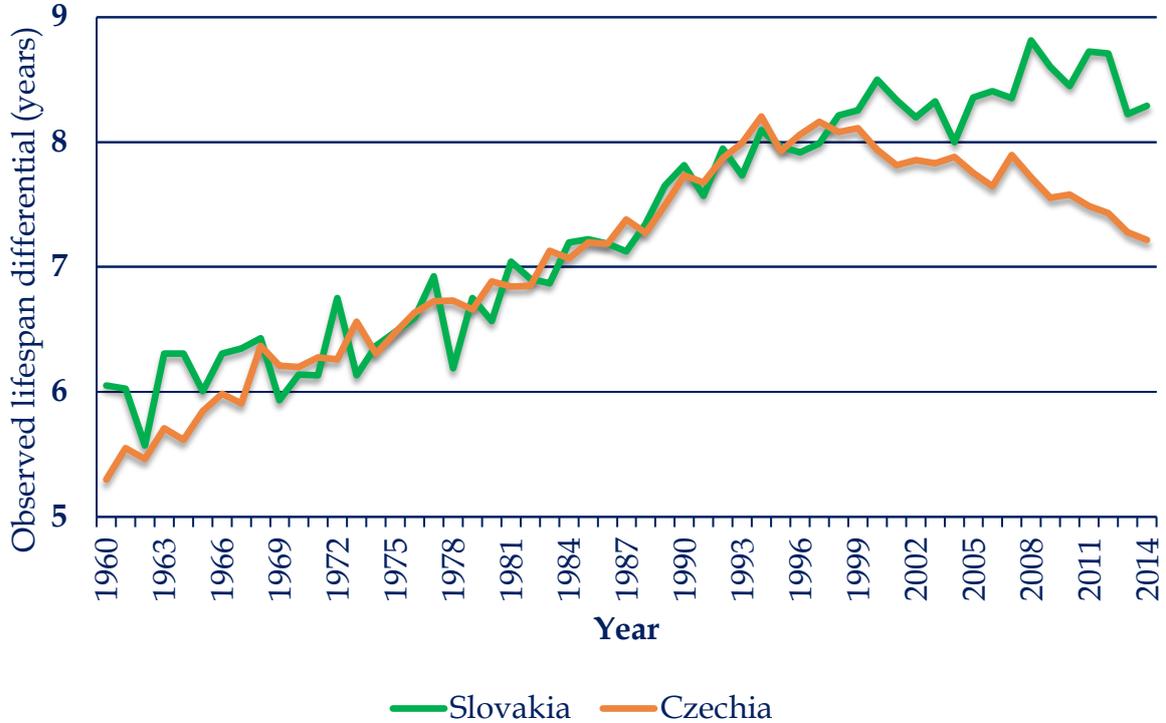

Figure 5: Dissolution of states - Slovakia and Czechia

*Precision of approximation for lifespan differential from subpopulations*

In this subsection we perform numerical testing of the approximation scheme for a population consisting of two non-overlapping subpopulations, proposed in section Methods and data. Namely, we would like to show that using $\Delta l_{app}^{tot}(t)$ as an approximation for $\Delta l^{tot}(t)$ is numerically very precise. To demonstrate the quality of this approximation scheme, we construct five synthetic populations from the studied dataset. In particular, we choose five pairs of countries and for each pair we create a synthetic population by simply adding corresponding numbers. For example, the number of male deaths in age range from 65 to 66 years for the synthetic population is obtained by summing number of male deaths in the age range from 65 to 66 years for the country 1 and the number of male deaths in the age range from 65 to 66 years for the country 2. In this way, from the data for the synthetic population we can calculate $\Delta l^{tot}(t)$, whereas from the data for country 1 and country 2 we can calculate terms in Eq. (7) and Eq. (8), $\Delta l_{app}^{tot}(t)$ and $\Delta l_{corr}^{tot}(t)$. In Table 1 we give the accuracy of the proposed approximation scheme given by the ratio of the correction term $\Delta l_{corr}^{tot}(t)$ and the total result $\Delta l^{tot}(t)$. From the Table 1 we can see that the approximation $\Delta l^{tot}(t) \cong \Delta l_{app}^{tot}(t)$ is accurate up to several percent.

Table 1: Precision analysis of the approximation for the lifespan differential.

| Country 1 | Russia | Belgium | Sweden | France | Portugal |
|---|---|---|---|---|---|
| Country 2 | Ukraine | Netherlands | Italy | Slovakia | Belarus |
| Years | 1960-2013 | 1960-2012 | 1960-2012 | 1960-2014 | 1960-2012 |
| $\max \dfrac{|\Delta l^{tot}_{corr}(t)|}{\Delta l^{tot}(t)}$ (%) | 1.11 | 0.39 | 0.87 | 0.53 | 2.32 |

*Lifespan differential for countries with significant immigration – Israel from 1990 to 2000*

Our final case study is the evolution of the observed lifespan differential in Israel which is depicted in Figure 6. A prominent feature of the observed lifespan differential dynamics is a considerable surge in the lifespan differential value starting approximately from 1990. Indeed, whereas in the period prior to 1990 the value of the lifespan differential oscillates around a stable value with no noticeable trend, in the period from 1990 to 2000 the value of the observed lifespan differential for Israel increased for a factor of almost two.

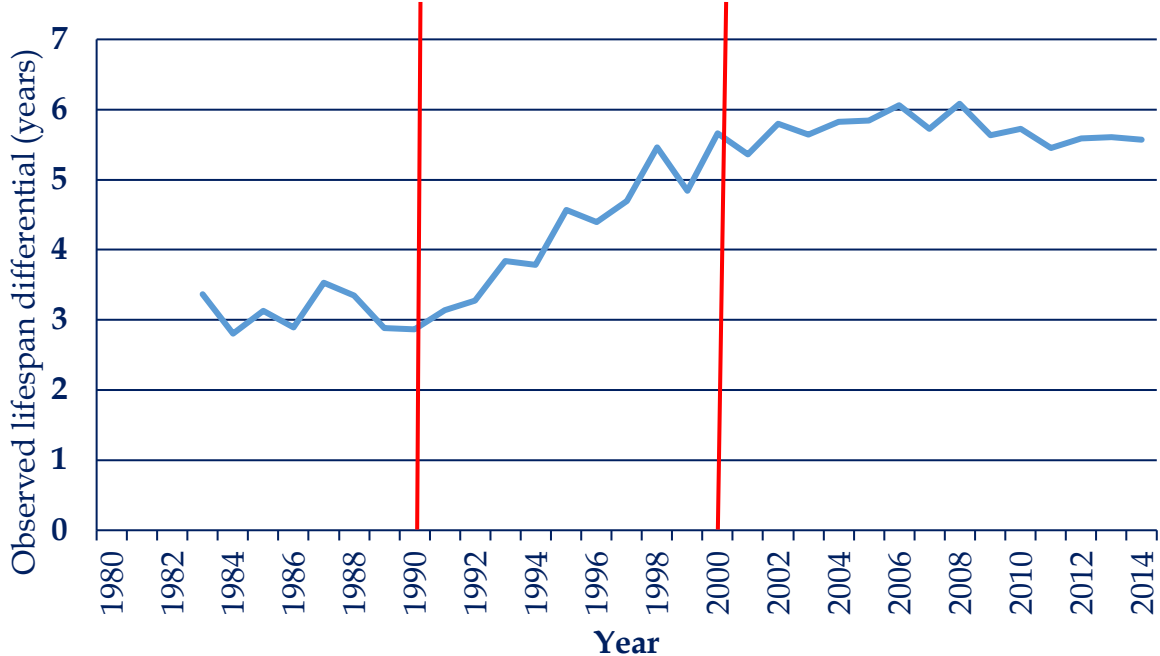

Figure 6: Dynamics of observed lifespan differential in Israel 1990 - 2000

On the other hand, in the decade starting in 1990 Israel experienced an immigration wave, with large majority of immigrants coming from the republics of the former Soviet Union (FSU). The natural idea of studying the interconnection of mortality patterns and immigration in Israel has already been presented in (Raphael, 2016). The aim of this subsection is to provide approximate quantitative understanding of the specific observed lifespan differential dynamics for Israel presented in Figure 6.

The main assumption of our approach is that the immigrant population maintains mortality patterns (in particular the observed lifespan differential) of its native country, i.e. the country of emigration, for some time after immigration. In a way it is natural that the immigrant population also imports its mortality patterns, at least in short time after immigration. The persistence of these patterns is dependent on a number of factors acting in opposite directions. Namely, immigrants to a more developed country usually benefit from better healthcare, working and living conditions and better societal organization in general. On the other hand, immigrants face increased stress of adaptation to novel circumstances, cultural change and disruption of personal relationships. It is not a priori clear which of these competing factors should outweigh in the dynamics of the observed lifespan.

As the observed lifespan differential has its highest values for FSU republics (among the studied countries), whereas the value of the lifespan differential for Israel prior 1990 was among the lowest for the studied countries, immigration of FSU citizens in yearly levels of several percent of Israeli population could be sufficient to produce a sizeable increase in the lifespan differential.

In order to build quantitative understanding of the lifespan differential in Israel in the period from 1990 till 2000, we develop a model relying on the following assumptions:

- the population of Israel is divided into a subpopulation of immigrants from FSU (subpopulation 1) and the rest of population (subpopulation 2);
- in the studied period both subpopulations maintain its total mortality rates and lifespan differentials at the level of 1990;
- the observed lifespan differential for the total Israeli population can be calculated using the first term in the expression (6), i.e. using $\Delta l_{app}^{tot}(t)$, from the data for subpopulations 1 and 2. The acceptability of this approximation follows from the numerical analysis in the previous subsection;
- total number of females dying in some year in the expression (7) can be replaced by the total number of deaths. This amounts to assumption that the proportion of the number of female deaths in the total number of deaths is the same for both subpopulation 1 and subpopulation 2.

In our calculations we use the data on number of immigrants from FSU from www.jewishvirtuallibrary.org, given in the second column of Table 2. In the third column of Table 2 we also calculate the cumulative number of immigrants starting from year 1990 and in the fourth column we present the data on total population of Israel taken from the Human Mortality Database. In the fifth column we give calculated values of the lifespan differential using the expression (7), whereas in the last column there are numbers for the observed lifespan differential for Israel.

Table 2: The observed and calculated lifespan differential for Israel from 1990 to 2000.

| Year | $N_{imm}$ | $N_{imm}^{cum}$ | $N_{pop}$ | $\Delta l_{calc}^{Israel}$ | $\Delta l^{Israel}$ |
|---|---|---|---|---|---|
| 1990 | 185227 | 185227 | 4821735 | 3.453912 | 3.139568 |
| 1991 | 147839 | 333066 | 5058838 | 3.859149 | 3.278163 |
| 1992 | 65093 | 398159 | 5195918 | 4.014542 | 3.843507 |
| 1993 | 66145 | 464304 | 5327647 | 4.16391 | 3.783342 |
| 1994 | 68079 | 532383 | 5471545 | 4.306062 | 4.568509 |
| 1995 | 64848 | 597231 | 5618981 | 4.430358 | 4.390132 |
| 1996 | 59048 | 656279 | 5757873 | 4.53557 | 4.69952 |
| 1997 | 54621 | 710900 | 5899932 | 4.623851 | 5.456505 |
| 1998 | 46032 | 756932 | 6041405 | 4.688403 | 4.836335 |
| 1999 | 66848 | 823780 | 6209145 | 4.786938 | 5.659632 |
| 2000 | 50817 | 874597 | 6369266 | 4.848443 | 5.359469 |

Source: Immigration to Israel: Total Immigration, by Country per Year. https://www.jewishvirtuallibrary.org/total-immigration-to-israel-by-country-per-year

The calculated value of the lifespan differential, $\Delta l_{calc}^{Israel}$, is obtained using (7) and approximations specified above:

$$\Delta l_{calc}^{Israel} = \frac{N_{imm}^{cum} m_{imm} \Delta l^{FSU} + (N_{pop} - N_{imm}^{cum}) m_{Israel,1990} \Delta l_{pre\ 1990}^{Israel}}{N_{imm}^{cum} m_{imm} + (N_{pop} - N_{imm}^{cum}) m_{Israel,1990}}. \quad (9)$$

As the precise data on number of immigrants from individual republics of FSU have not been available to us, we have used the information that 32% of all immigrants came from Russia, 33% percent from Ukraine, 8% from Belarus, 2% from Baltic republics (Estonia, Latvia, Lithuania) and the rest from other FSU republics. As Human Mortality Database only provides data for these six specified FSU republics, we approximated all mortality properties for FSU with a weighted average of properties of these six republics with the following weights: Russia (0.32), Ukraine (0.33), Belarus (0.08), Estonia (0.02/3), Latvia (0.02/3) and Lithuania (0.02/3) (Source: http://www.cbs.gov.il/statistical/immigration_e.pdf). This approach gives us the following indicators needed in (9): $\Delta l^{FSU}$=12.18649, $\Delta l_{pre\ 1990}^{Israel}$=2.86, $m_{imm}$=11.5764444‰, $m_{Israel,1990}$=6.8‰.

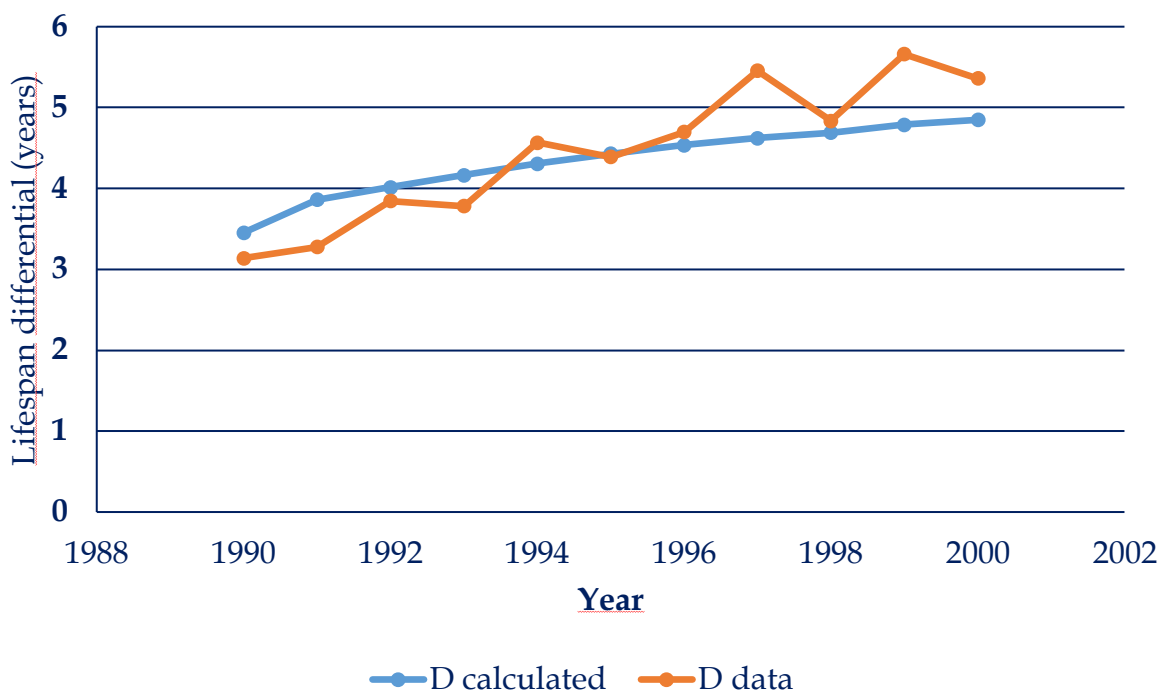

Figure 7: Calculated vs. observed dynamics of lifespan differential in Israel 1990 - 2000

All approximations made are justified since we do not have the data on mortality of males and females of the immigrant population after immigration. However, even an approach based on these approximations succeeds in explaining a large part of the lifespan differential growth in Israel from 1990 to 2000.

It is reasonable to expect that the calculations with more precise data would bring the measured and calculated values of the lifespan differential for Israel to even better agreement.

It is further intriguing to contemplate the deviation of calculated lifespan differential (based on the assumptions of imported mortality characteristics) from the observed one with time since immigration as a measure of integration of immigrants in the society of the host country. Clearly, such an application of our method requires more detailed data on immigration and mortality of immigrants.

**Discussion and directions for further research**

The identified trends of the observed lifespan differential for various countries in the studied sample indicate differing rates of increase of the observed average lifespan for male and female population. For a population exhibiting a generic trend of growing observed lifespan differential, followed by the declining (or stagnating) observed lifespan differential, the average observed lifespan of female population grows faster that the average observed lifespan of the male population in the first phase. In the second phase the change of the average observed lifespan in time of the male population surpasses (or approximately equals) the change of the average observed lifespan in time of the female population. It is possible to envisage several processes that might contribute to the observed dynamics of lifespan differential, in particular to the reversal of the growing trend for the lifespan differential. Each one of them requires a dedicated study and it is currently a hypothesis left for future work. These processes have already been considered as sources of sex differences in life expectancy or mortality rates (e.g. Waldron 1993; Case and Paxson 2005; Preston and Wang 2006; Oksuzyan et al. 2008; Rogers et al. 2010) and it also reasonable to consider them as sources of sex differences in the observed lifespan. These hypotheses comprise:

- Deindustrialization and transition to postindustrial societies increase lifespan of men more than lifespan of women. With deindustrialization and the transition to postindustrial society the percentage of workforce working physically diminishes. The incidence of physical work-related disorders and the general level of work induced attrition is considerably reduced. As the majority of jobs requiring physical work have been historically occupied by male workforce, this systemic change has affected male population more than the female population.

- Changes in social rights and pension systems decrease the lifespan of women more than lifespan of men. Regulatory measures aimed at achieving equality of female and male workforce have increased average working hours and age of retirement for female population. Larger exposure of female workforce to these working conditions may have reduced the average observed lifespan growth rate of female population and to the corresponding reduction of the observed lifespan differential.

- In recent years, more efficient healthcare increases lifespan of men more than lifespan of women. Male population is practicing increasingly healthier lifestyle, including more effective use of available healthcare and reduction in unhealthy habits such as smoking, excessive alcohol consumption or unhealthy diet. This process would increase average lifespan of male population and correspondingly reduce the observed lifespan differential.

**Conclusions**

Historical records of growing trends for various longevity measures such as the average observed lifespan or lifetime expectancy at first impose the conclusion of near monotonicity of these trends. However, when the difference in the average observed lifespan between females and males is studied, there is evidence across many countries in the world that the observed lifespan differential, as defined in this paper, changes trend from increasing to decreasing, or at least stagnating. Some causes for such a change may be country specific, but universal trends such as larger participation of women in workforce, changes in pension systems, accessibility of healthcare and adoption of healthier lifestyles all seem as plausible contributors to such a change. The effort of investigating their role in the observed lifespan differential dynamics appears worthwhile. The available data on mortality trends also provide several interesting case studies of strong association of the observed lifespan differential dynamics and disruptive effects such as wars, policy campaigns, dissolutions of states or integration of states on the supranational level. These associations may serve as a solid starting point for research how the said factor influence the trends of the observed lifespan differential. Finally, a method for calculating the observed lifespan differential of a population from the observed lifespan differentials of its subpopulations has been introduced. The case of lifespan differential for Israel in the period 1990-2000 shows that this method can be helpful in explaining the change of the observed lifespan differential in circumstances of significant immigration. In this time of large migrations, such a method is a welcome addition to existing measures and methods of studying how migrations affect all aspects of society.

**Acknowledgements**

The research of the authors (T.Ć., R. M. and H. Š.) presented in this paper was funded by the project *Computer model of demographic dynamics in the Republic of Croatia – quantitative foundation for future demographis policies* of the Adris foundation. This research has also been supported by the European Regional Development Fund under the grant KK.01.1.1.01.0009 (DATACROSS) (H. Š.). The authors would like to thank Alison van Raalte for useful comments on the manuscript.

**Appendix**

In this appendix we present five year moving average of the observed lifespan differential for all countries with available data in Human Mortality Database. All plots use the same scales for x-axis (1960- 2014) and y-axis (2-14 years) for reasons of comparability.

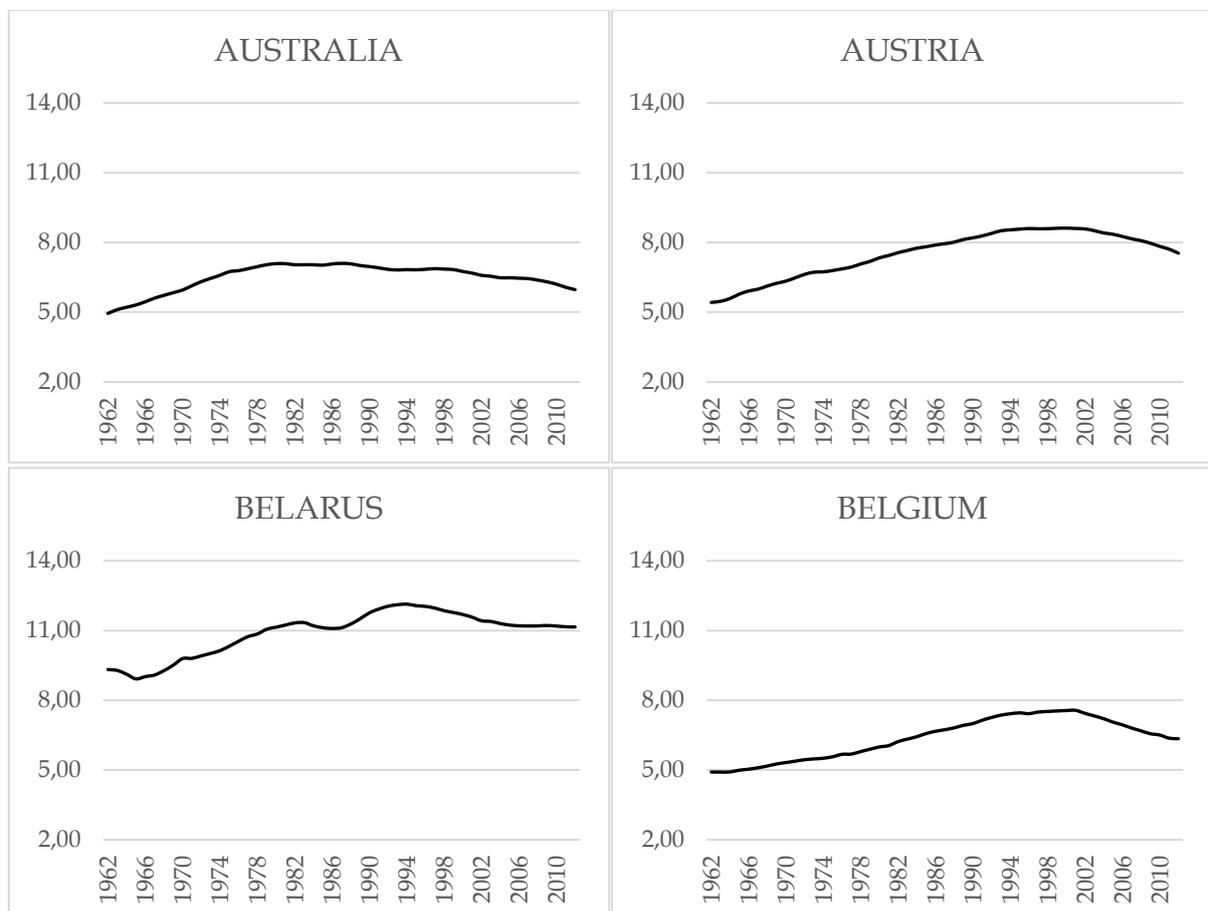

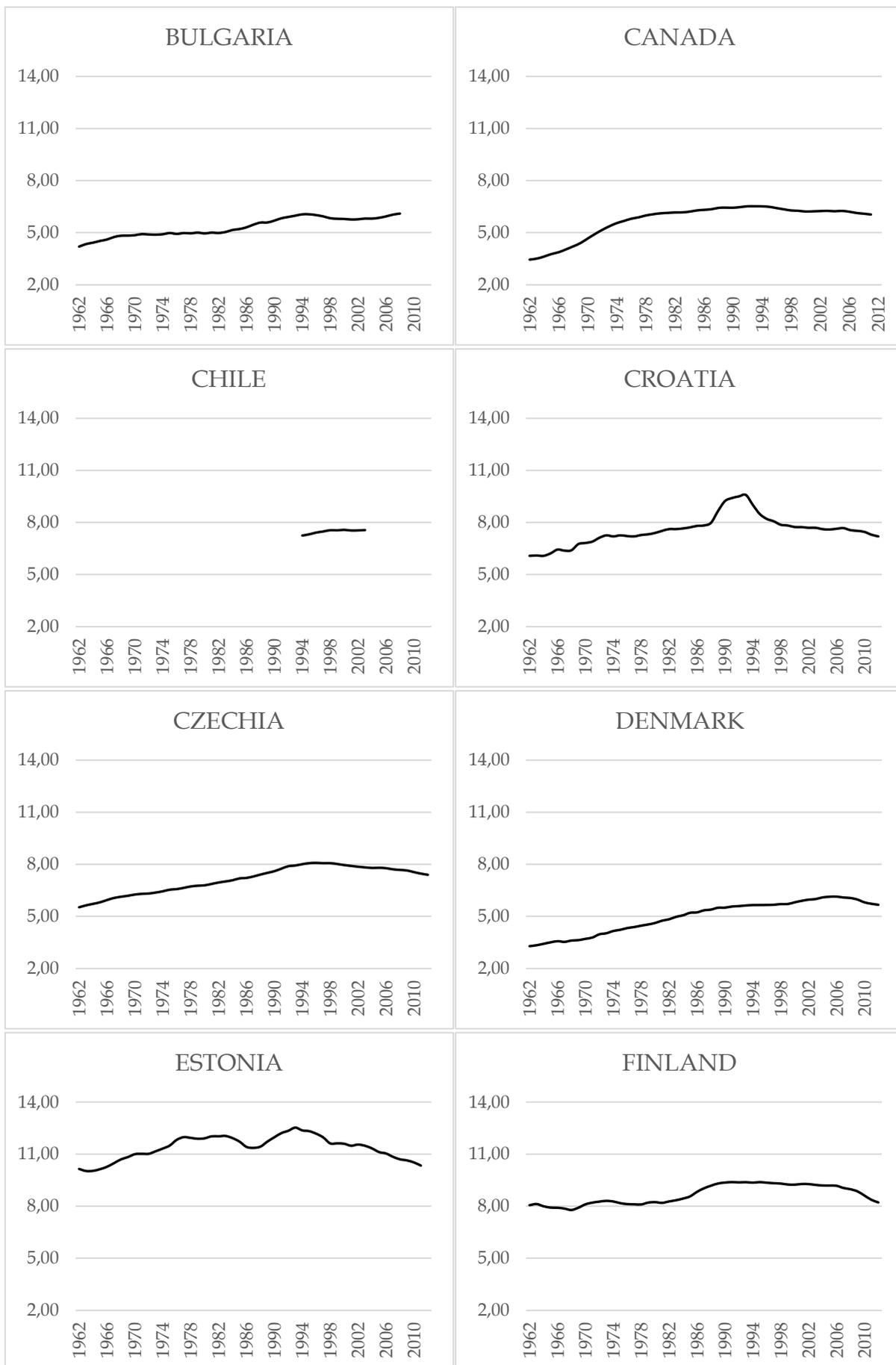

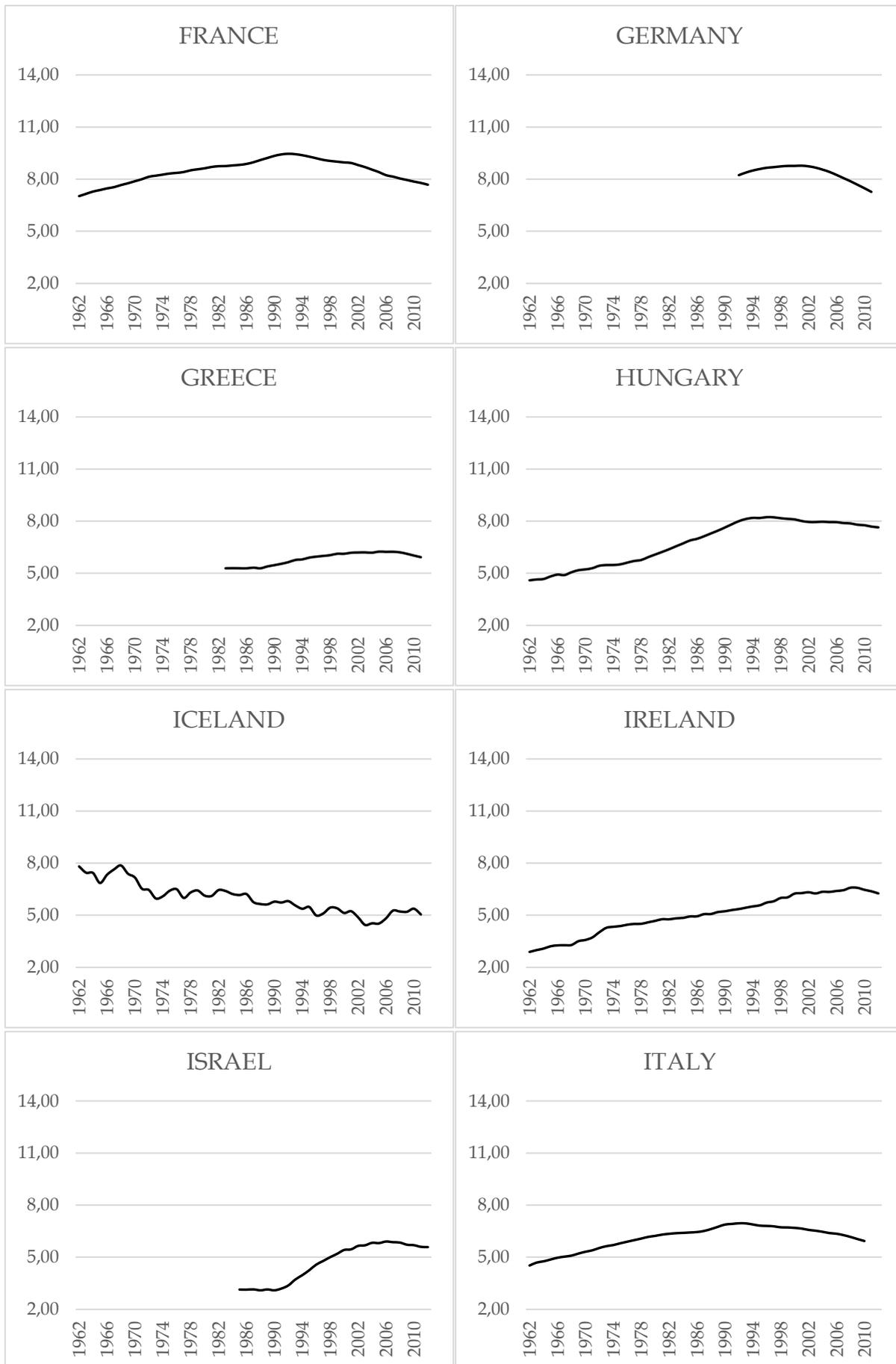

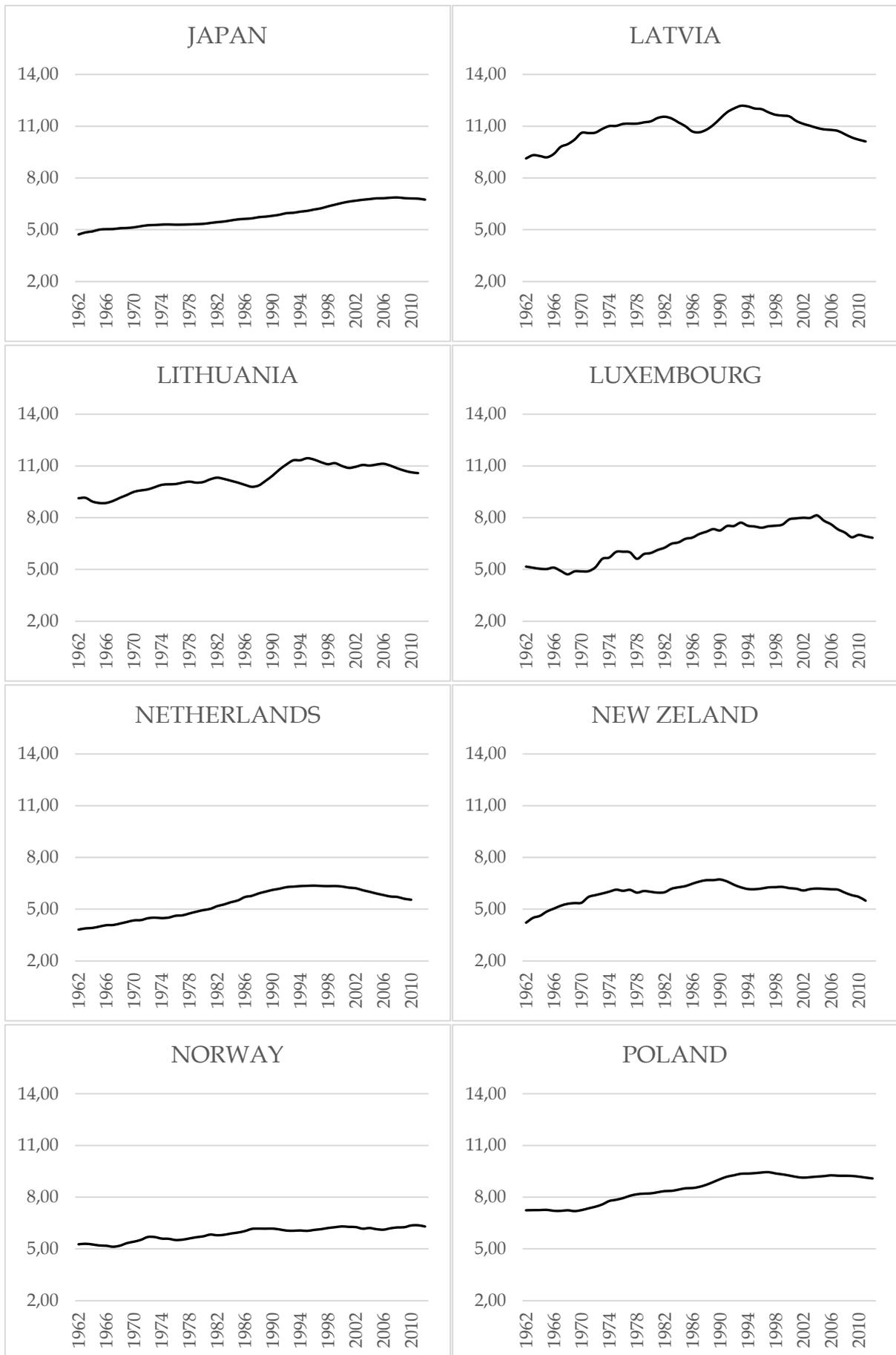

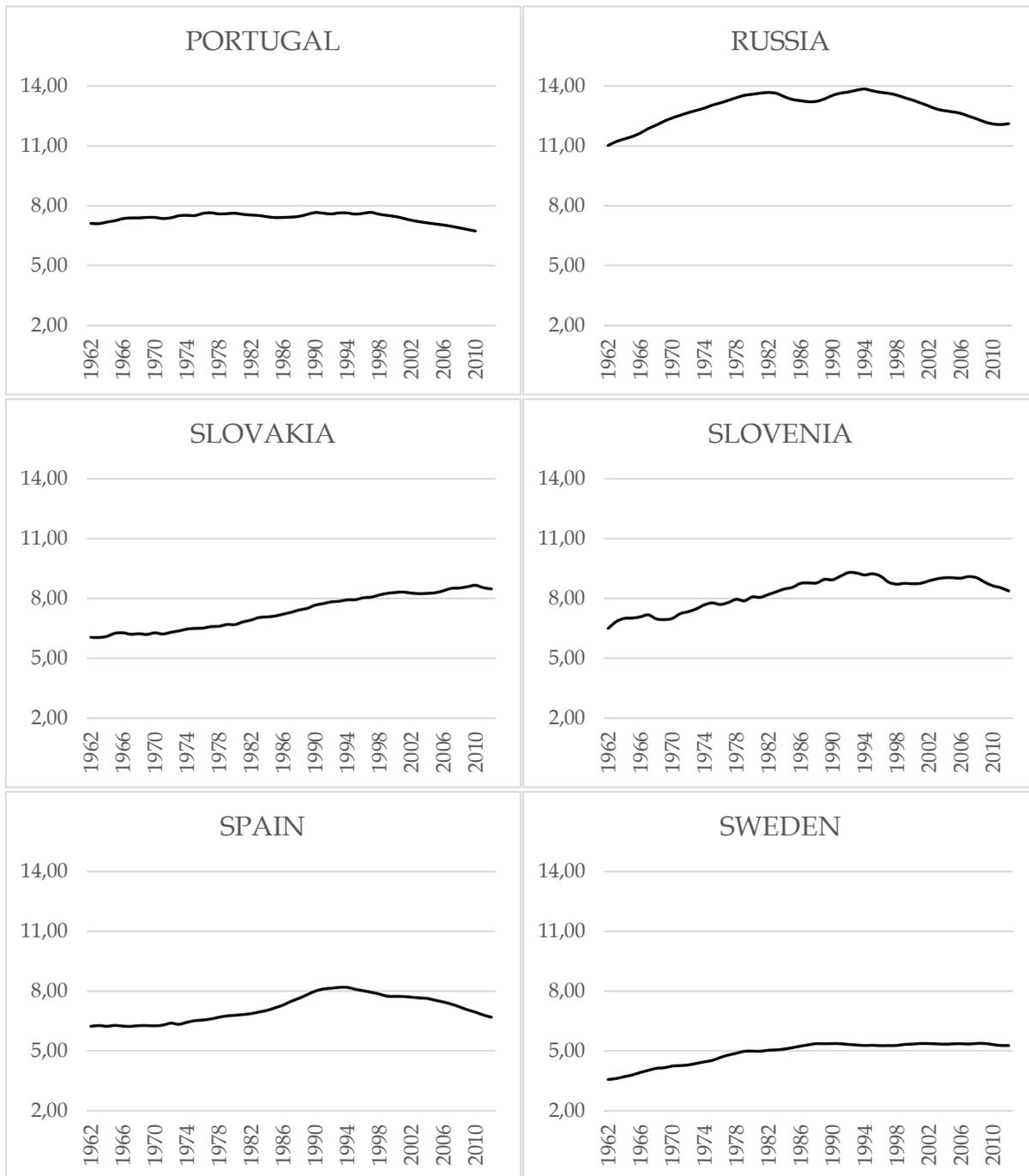

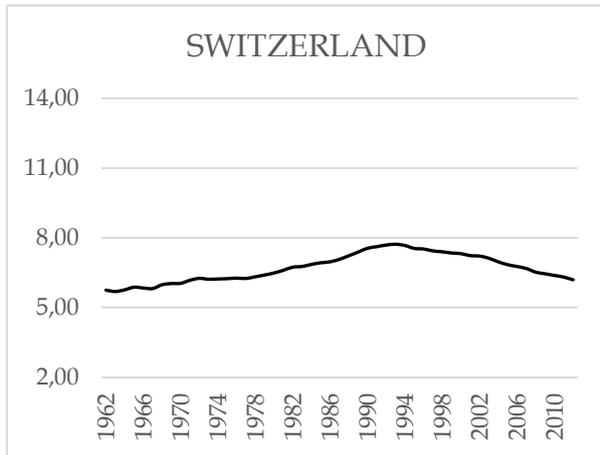
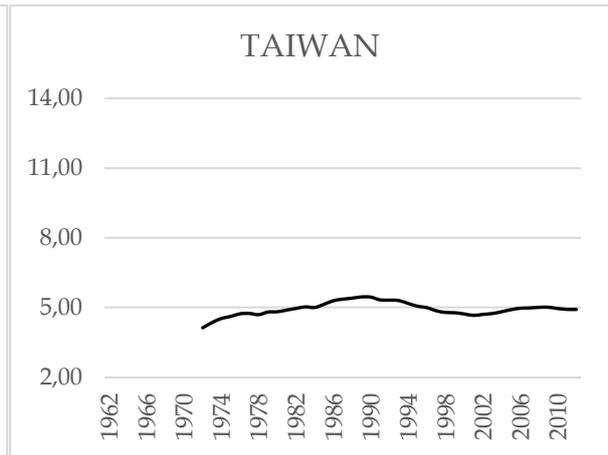
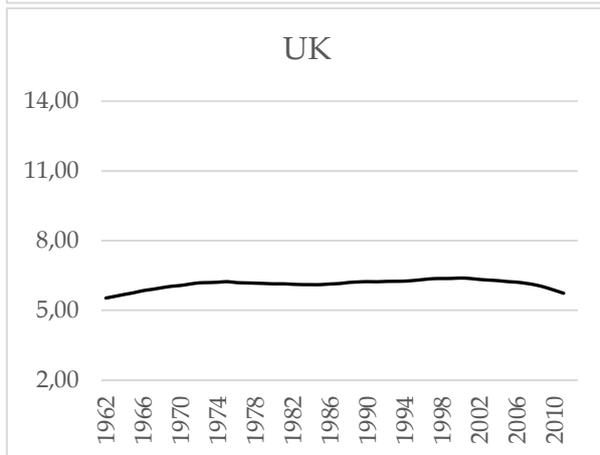
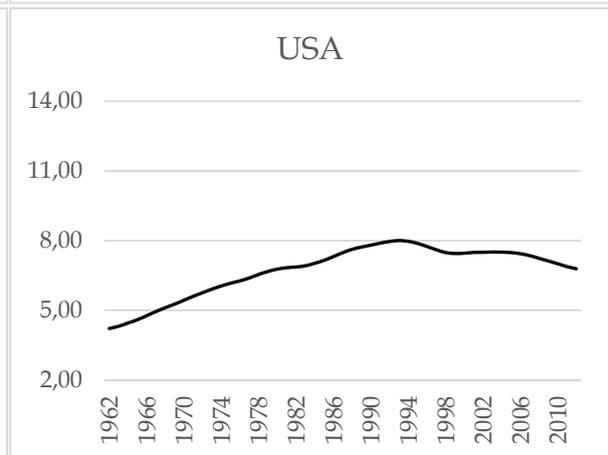
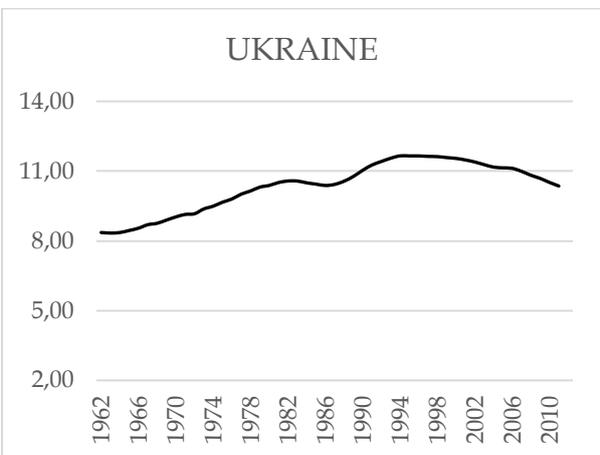